\documentclass[conference]{IEEEtran}

\usepackage{amsmath,amssymb,bm,mathtools}
\usepackage{graphicx}
\usepackage{booktabs}
\usepackage{array}
\usepackage{cite}
\usepackage{xurl}
\IEEEoverridecommandlockouts
\newcommand{\Esub}{\mathrm{E}}
\newcommand{\Psub}{\mathrm{P}}

\graphicspath{{./}{figures_gm_singlecolumn_v2/}}
\begin{document}
\title{Multi-Timescale Deliverability of Shared Hybrid Storage for AI Data Center Campuses}

\author{
    Zihan Meng\textsuperscript{*},
    Ruizhang Yang\textsuperscript{\ensuremath{\ddag}},
    Chenxi Hu\textsuperscript{\ensuremath{\dag}}, and
    Francis Yunhe Hou\textsuperscript{*} \\ 
    \textit{\textsuperscript{*}Department of Electrical and Computer Engineering, The University of Hong Kong, Hong Kong}\\
    \textit{\textsuperscript{\ensuremath{\dag}}Department of Computer Science, The Hang Seng University of Hong Kong, Hong Kong}\\
    \textit{\textsuperscript{\ensuremath{\ddag}}Computer, Electrical and Mathematical Sciences and Engineering (CEMSE) Division,}\\
    \textit{King Abdullah University of Science and Technology (KAUST), Thuwal, Saudi Arabia}
    
    \thanks{This work was supported by the National Natural Science Foundation of China (NSFC) under Grant 52437005.}
}

\maketitle

\begin{abstract}
Artificial intelligence data center (AIDC) campuses contain sustained power variations and subsecond power
fluctuations, requiring both sufficient stored energy and fast response. A shared hybrid energy storage system (HESS) can mitigate aggregate AIDC power deviations, but interval-based schedules cannot capture the delay between commanded and delivered power at the campus point of common coupling (PCC). To this end, this paper proposes a multi-timescale scheduling model for mitigating AIDC power fluctuations with the shared HESS. First, it models electrical pooling at the campus bus and constrains the shared HESS to a single PCC injection. Second, a two-layer model passes five-minute scheduled power, energy state, and reserve capacity to a causal fast response model that distinguishes commanded power from delivered power. Case studies using the National Laboratory of the Rockies (NLR) GenAI Power Profiles Dataset show reductions in both PCC and fast integral absolute error (IAE), while retaining most of the improvement available under ideal response.

\end{abstract}


\section{Introduction}
\label{sec:introduction}

\IEEEPARstart{A}{rtificial} intelligence is turning data centers into some of
the largest loads connected to power systems. Global data-center
electricity consumption is projected to increase from about 415~TWh in 2024
to 945~TWh in 2030 \cite{IEA2025EnergyAI}. Artificial intelligence data
centers (AIDCs) are increasingly built as large campuses connected through
one point of common coupling (PCC). Their growth increases both electricity
demand and the power variation
\cite{NERC2025LargeLoads}.

AIDC demand is challenging because its magnitude and rate of change matter on
different time scales. Both training and inference workloads can produce sustained demand and
subsecond power changes.
Measurements of H100 systems and public facility profiles show that subsecond
ramps coexist with sustained power variations
\cite{Latif2025Power, Vercellino2026}. The former requires rapid power
response, and the latter requires sufficient stored energy. Meeting both
requirements efficiently is difficult for one storage technology. 

Energy storage can regulate these variations without rescheduling computing tasks. In a Norwegian field trial, uninterruptible
power supply (UPS) batteries responded to a frequency event without
interrupting the protected load \cite{Takci2025Flexibility}. This application
demonstrates that storage can support grid operation while maintaining
data-center continuity.

However, a single type of storage cannot efficiently provide both the
energy required for sustained AIDC variations and the rapid response required for subsecond changes. Existing studies have applied HESSs to coordinated dispatch and power fluctuation mitigation
\cite{Zheng2017HESS, Wang2023Hierarchical, You2026SourceSide, Ko2025Mitigation}.
They show the benefit of matching storage characteristics to AIDC power
variations. The campus problem additionally requires distinguishing one
shared PCC injection from independent local actions and checking its
delivery against the aggregate demand.

A shared HESS serves the net campus requirement with common energy and
fast response capability. Current studies
allocate shared storage capacity among users over scheduling intervals
\cite{Han2023Shared, Bian2024Shared, Zhu2021Distributed}, but do
not establish whether one HESS can deliver subsecond response. At a campus PCC, AIDC residuals combine electrically and the shared HESS
provides one shared PCC injection rather than separate facility actions. Its power is not delivered instantaneously
\cite{DOE2024Latency}. Hence, the physical deliverability of
shared HESS operation remains unresolved.

\begin{figure}[t]
\centering
\includegraphics[width=\columnwidth]{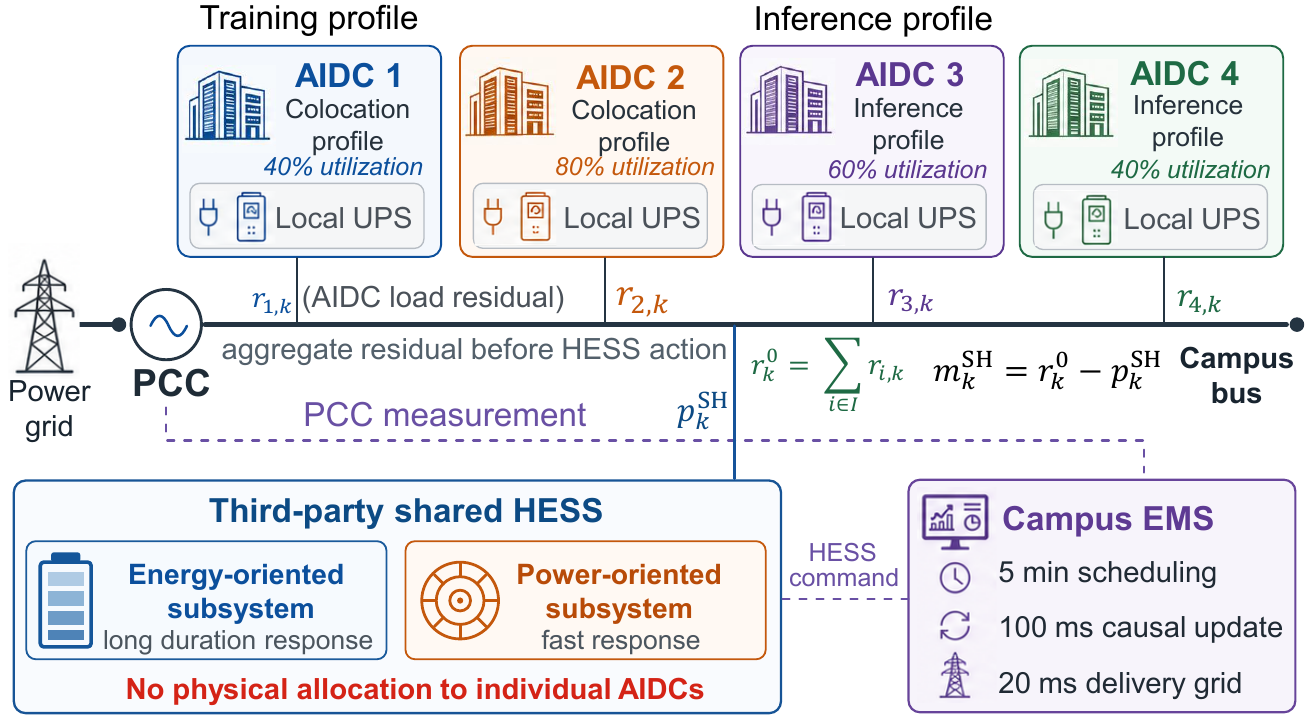}
\caption{Campus shared HESS architecture and service boundary.}
\label{fig:physical_architecture}
\end{figure}

To address this gap, this paper develops an AIDC campus model that connects
the HESS schedule at the PCC with the fast power changes of AIDC workloads.
It models one shared HESS serving the combined campus demand and evaluates
the power it can actually deliver when storage limits and response delay
are considered. The main contributions are as follows:
\begin{enumerate}
    \item A physical campus PCC model is developed for one shared HESS. It captures the electrical pooling of AIDC residuals before storage acts and restricts the shared HESS to one PCC injection instead of virtual allocations to individual AIDCs.

    \item A mixed-integer linear programming (MILP) model couples minute-level scheduling with subsecond power delivery. The slow scheduling layer prepares the power, stored energy, and reserve of the shared HESS. The fast response layer distinguishes commanded and delivered power using causal measurements, and checks delivery under response delays and storage operating limits.

    \item The model is validated using the National Laboratory of the Rockies (NLR) GenAI Power Profiles Dataset. Capacity-matched local and BESS cases evaluate the shared connection and hybrid composition, while the ideal response case quantifies the effect of response delay. The results demonstrate effective regulation over both time scales and improved fast response over the capacity-matched BESS.
\end{enumerate}

\section{Shared HESS Configuration and Slow Scheduling}
\label{sec:campus_model}

\subsection{Campus Configuration and Load Aggregation}

The AIDC campus architecture is shown in Fig.~\ref{fig:physical_architecture}. Local UPSs protect server continuity, while the shared HESS changes net
utility exchange through the PCC. The slow scheduling layer determines
minute-level power, stored energy, and reserve; the fast response layer
evaluates subsecond delivery under controller and device delays.

Let $\mathcal I$ index the AIDCs and
$\mathcal D=\{\Esub,\Psub\}$ the shared HESS subsystems. The energy-oriented
subsystem $\Esub$ provides longer-duration support, while the power-oriented
subsystem $\Psub$ provides faster response. Both training and inference can
contain sustained and subsecond variations, so HESS response is assigned
by time scale. Five-minute intervals are indexed by
$\mathcal K=\{0,\ldots,K-1\}$, with $\Delta T=5/60$~h. For AIDC $i$, the
residual relative to the causal reference $\bar L_{i,k}$ is
\begin{equation}
r_{i,k}=L_{i,k}-\bar L_{i,k},\qquad
r_k^{0}=\sum_{i\in\mathcal I}r_{i,k}.
\label{eq:campus_residual}
\end{equation}
The pre-HESS residual $r_k^0$ reflects electrical pooling at the common
bus. Opposite residuals can cancel without storage action in both shared
and local configurations. The configurations differ in their storage
control, not in electrical pooling.
\subsection{Slow Scheduling Layer and Reserve Constraints}

The slow scheduling layer defines the centralized operating principle. A
campus energy management system schedules the shared HESS, with positive power denoting discharge to the campus bus. For $d\in\mathcal D$, the slow power variables satisfy
\begin{equation}
\begin{aligned}
p_{d,k}&=p_{d,k}^{\mathrm{dis}}-p_{d,k}^{\mathrm{ch}},\\
0\leq p_{d,k}^{\mathrm{dis}}&\leq\bar P_du_k,\qquad
0\leq p_{d,k}^{\mathrm{ch}}\leq\bar P_d(1-u_k).
\end{aligned}
\label{eq:slow_power}
\end{equation}
The signed output $p_{d,k}$ is the contribution of subsystem $d$ to the campus PCC. The common binary variable $u_k\in\{0,1\}$ implements an EMS operating
policy that gives both subsystems the same direction and excludes internal
counterflow. Their stored energy satisfies
\begin{equation}
\begin{aligned}
e_{d,k+1}={}&e_{d,k}+\eta_d^{\mathrm{ch}}p_{d,k}^{\mathrm{ch}}\Delta T
-\frac{p_{d,k}^{\mathrm{dis}}\Delta T}{\eta_d^{\mathrm{dis}}},\\
s_d^{\min}\bar E_d\leq{}&e_{d,k}\leq s_d^{\max}\bar E_d,
\qquad k=0,\ldots,K,\\
e_{d,0}={}&e_d^{\mathrm{ini}},\qquad
|e_{d,K}-e_d^{\mathrm{ini}}|\leq\varepsilon_d^{\mathrm{ter}}.
\end{aligned}
\label{eq:slow_energy}
\end{equation}
The stored energy $e_{d,k}$ carries the consequence of the scheduled injection across slow intervals and then determines the physical availability of the sustained service. Its terminal state prevents end-of-day depletion from being counted as a free improvement.

The reserve passed to the fast response layer must be deliverable in both
power and energy,
\begin{equation}
\begin{aligned}
p_{d,k}^{\mathrm{dis}}+R_{d,k}^{\uparrow}\leq{}&\bar P_d,\qquad
e_{d,k}-s_d^{\min}\bar E_d\geq
\frac{R_{d,k}^{\uparrow}T_d^{\mathrm{res}}}
{3600\eta_d^{\mathrm{dis}}},\\
p_{d,k}^{\mathrm{ch}}+R_{d,k}^{\downarrow}\leq{}&\bar P_d,\qquad
s_d^{\max}\bar E_d-e_{d,k}\geq
\frac{\eta_d^{\mathrm{ch}}R_{d,k}^{\downarrow}T_d^{\mathrm{res}}}{3600}.
\end{aligned}
\label{eq:sustainable_reserve}
\end{equation}
The nonnegative reserves $R_{d,k}^{\uparrow,\downarrow}$ provide power
headroom and energy margins for a reserve duration $T_d^{\mathrm{res}}$
in seconds. The fast response layer additionally checks the total
commanded and delivered powers and the associated energy excursion.

\subsection{Shared and Local Storage Configurations}

The shared and local setting allow different storage actions even
when their total power and energy ratings are identical. Their PCC
balances are 
\begin{equation}
\begin{aligned}
p_k^{\mathrm{SH}}&=\sum_{d\in\mathcal D}p_{d,k}, &
m_k^{\mathrm{SH}}&=r_k^0-p_k^{\mathrm{SH}},\\
p_{i,k}^{\mathrm{loc}}&=\sum_{d\in\mathcal D}p_{i,d,k}^{\mathrm{loc}}, &
m_k^{\mathrm{loc}}&=\sum_{i\in\mathcal I}
\bigl(r_{i,k}-p_{i,k}^{\mathrm{loc}}\bigr).
\end{aligned}
\label{eq:topology_balances}
\end{equation}
The shared HESS provides one controllable injection at the PCC, so $p_k^{\mathrm{SH}}$ has no AIDC index. In the local configuration,
$p_{i,k}^{\mathrm{loc}}$ acts before campus aggregation, allowing each local
HESS to respond independently. The utility observes only their combined
effect, $m_k^{\mathrm{loc}}$. This differs from the installed capacity because the capacity-matched local benchmark satisfies
\begin{equation}
\sum_{i\in\mathcal I}\bar P_{i,d}^{\mathrm{loc}}=\bar P_d,
\qquad
\sum_{i\in\mathcal I}\bar E_{i,d}^{\mathrm{loc}}=\bar E_d,
\quad d\in\mathcal D.
\label{eq:capacity_matching}
\end{equation}
The capacity equalities in \eqref{eq:capacity_matching} give the shared and
distributed models identical campus-total MW and MWh ratings for each
functional subsystem. Each local port obeys the power, energy, and reserve constraints in
\eqref{eq:slow_power}--\eqref{eq:sustainable_reserve} using its own ratings
and direction mode. Thus, internal counterflow remains prohibited within one local HESS, while different AIDC ports may legitimately operate in opposite directions.

\section{Fast Response Modeling and Multi-Timescale Optimization}
\label{sec:fast_model}

\subsection{Fast Response Layer With Device Delays}

The fast response layer evaluates a window $w\in\mathcal W$ anchored to slow interval $k_w$. It inherits $p_{d,k_w}$, $e_{d,k_w}$, and
$R_{d,k_w}^{\uparrow,\downarrow}$ from the slow scheduling layer. At
controller update $q$, separated by $\Delta t_c$, the AIDC fast residual
$\rho_{i,w}(t)$ is sampled as
$\rho_{i,w,q}=\rho_{i,w}(q\Delta t_c)$. These residuals aggregate at the
campus PCC and are modeled by
\begin{subequations}
\label{eq:causal_split}
\begin{align}
\rho_w(t)&=\sum_{i\in\mathcal I}\rho_{i,w}(t),\qquad
\rho_{w,q}=\sum_{i\in\mathcal I}\rho_{i,w,q},
\label{eq:fast_aggregation}\\
\ell_{w,q}&=\alpha_s\ell_{w,q-1}+(1-\alpha_s)\rho_{w,q},
\  \alpha_s=e^{-\Delta t_c/\tau_s},
\label{eq:causal_filter}\\
\Delta\hat p_{\Esub,w,q}
&=\mathcal C(\ell_{w,q},\rho_{w,q}),\ 
\Delta\hat p_{\Psub,w,q}
=\rho_{w,q}-\Delta\hat p_{\Esub,w,q}.
\label{eq:power_split}
\end{align}
\end{subequations}
The causal state $\ell_{w,q}$, initialized by $\ell_{w,-1}=0$, extracts the
low-frequency component with time constant $\tau_s$. The mapping
$\mathcal C(a,b)=\operatorname{sgn}(b)\min(|a|,|b|)$ for $ab>0$, and zero
otherwise, assigns this component to the energy-oriented subsystem without
producing an opposing output. The power-oriented subsystem receives the
remainder. The commanded power
$p_{d,w,q}^{\mathrm{cmd}}=p_{d,k_w}+\Delta\hat p_{d,w,q}$ is held until the
next controller update, forming $p_{d,w}^{\mathrm{cmd}}(t)$ in
\eqref{eq:fopdt}. This causal split is fixed, not optimized and the MILP
selects a slow operating point compatible with its fast response.

The total dead time is
$D_d=T^{\mathrm{sen}}+T^{\mathrm{com}}+T^{\mathrm{cmp}}+T^{\mathrm{cmd}}
+T_d^{\mathrm{start}}$. The delivered power follows
\begin{equation}
\tau_d\dot x_{d,w}(t)+x_{d,w}(t)
=p_{d,w}^{\mathrm{cmd}}(t-D_d).
\label{eq:fopdt}
\end{equation}
The delivered power $x_{d,w}(t)$ starts from
$x_{d,w}(0)=p_{d,k_w}$. Its 95\% response time is
$T_{d,\mathrm{e2e}}^{95}=D_d-\tau_d\ln(0.05)$.

On the device grid $t_n=n\Delta\tau$, let
$d_d=\lceil D_d/\Delta\tau\rceil$ and
$\alpha_d=e^{-\Delta\tau/\tau_d}$. For $n\geq1$, the sampled response is
\begin{equation}
x_{d,w,n}=\alpha_dx_{d,w,n-1}
+(1-\alpha_d)\widetilde p_{d,w,n-d_d}^{\mathrm{cmd}}.
\label{eq:sampled_delivery}
\end{equation}
Here, $x_{d,w,0}=p_{d,k_w}$ and, for $n\geq0$,
$\widetilde p_{d,w,n}^{\mathrm{cmd}}
=p_{d,w,\lfloor n\Delta\tau/\Delta t_c\rfloor}^{\mathrm{cmd}}$.
For $n<0$, it equals $p_{d,k_w}$. Rounding $d_d$ upward prevents delivery
before the continuous dead time.

\subsection{Slow Scheduling With Fast Response Constraints}

The slow scheduling and fast response layers are coupled through the inherited scheduled power, stored energy state, and reserve capacity. With $\rho_{w,n}=\rho_w(t_n)$, both the commanded and delivered increments
must remain within the reserve scheduled at $k_w$,
\begin{equation}
\begin{aligned}
-R_{d,k_w}^{\downarrow}
&\leq p_{d,w,q}^{\mathrm{cmd}}-p_{d,k_w}
\leq R_{d,k_w}^{\uparrow},\\
-R_{d,k_w}^{\downarrow}
&\leq x_{d,w,n}-p_{d,k_w}
\leq R_{d,k_w}^{\uparrow},\\
m_{w,n}^{F}
&=\rho_{w,n}
-\sum_{d\in\mathcal D}(x_{d,w,n}-p_{d,k_w}).
\end{aligned}
\label{eq:fast_balance}
\end{equation}
The commanded power $p_{d,w,q}^{\mathrm{cmd}}$ is the controller output,
and $x_{d,w,n}$ is the power delivered through
\eqref{eq:sampled_delivery}. Only the delivered increment reduces the PCC fast
residual $m_{w,n}^{F}$. Both powers obey the form of \eqref{eq:slow_power}, with separate
command and delivery binary modes, each common to the two subsystems
and distinct from $u_k$. With
$x_{d,w,n}=x_{d,w,n}^{\mathrm{dis}}-x_{d,w,n}^{\mathrm{ch}}$, the corresponding
energy excursion satisfies
\begin{equation}
\begin{aligned}
\delta e_{d,w,n+1}^{F}
={}&\delta e_{d,w,n}^{F}
+\eta_d^{\mathrm{ch}}
\bigl(x_{d,w,n}^{\mathrm{ch}}-p_{d,k_w}^{\mathrm{ch}}\bigr)
\frac{\Delta\tau}{3600}\\
&-\frac{
\bigl(x_{d,w,n}^{\mathrm{dis}}-p_{d,k_w}^{\mathrm{dis}}\bigr)\Delta\tau}
{3600\eta_d^{\mathrm{dis}}},\\
\delta e_{d,w,0}^{F}
={}&0,\ 
s_d^{\min}\bar E_d
\leq e_{d,k_w}\!+\delta e_{d,w,n}^{F}
\leq s_d^{\max}\bar E_d.
\end{aligned}
\label{eq:fast_energy}
\end{equation}
The excursion $\delta e_{d,w,n}^{F}$ measures the additional energy relative to the slow power and avoids double-counting the five-minute schedule. Each fast window acts as a nested feasibility test, so the excursion is constrained around $e_{d,k_w}$ but is not passed to the next slow interval. Prior to the smallest $D_d$, the delivered increment is zero and the initial residual remains unserved.

The resulting MILP jointly represents the slow scheduling and fast
response layers,
\begin{equation}
\begin{aligned}
(\mathcal P^{\mathrm{SH}})\quad
\min_{\bm\Omega^S,\bm\Omega^F}\quad
J={}&\lambda_m\widetilde J_m
+\lambda_{\Delta}\widetilde J_{\Delta}
+\lambda_f\widetilde J_f\\[-1mm]
&+\lambda_c\widetilde J_c
+\lambda_R\widetilde J_R,\\
\mathrm{s.t.}\quad
&\bm\Omega^S\in\mathcal F^S,\quad
\bm\Omega^F\in\mathcal F^F(\bm\Omega^S).
\end{aligned}
\label{eq:optimization}
\end{equation}
The slow scheduling variables $\bm\Omega^S$ define the scheduled power, stored energy state, reserve capacity, direction, and PCC residual, while the fast response variables $\bm\Omega^F$ define the commanded and delivered power. The sets $\mathcal F^S$ and $\mathcal F^F$ collect
the slow constraints and the fast response, power, and energy constraints,
respectively. Here,
$J_m=\sum_k|m_k^{\mathrm{SH}}|\Delta T$,
$J_{\Delta}=\sum_{k\geq1}|m_k^{\mathrm{SH}}-m_{k-1}^{\mathrm{SH}}|$, and
$J_f=\sum_{w,n}|m_{w,n}^{F}|\Delta\tau$ measure PCC integral absolute error (IAE), PCC variation, and
fast IAE. The terms
$J_c=\sum_{d,k}c_d^{\mathrm{thr}}
(p_{d,k}^{\mathrm{ch}}+p_{d,k}^{\mathrm{dis}})\Delta T$
and $J_R=\sum_{d,k}(R_{d,k}^{\uparrow}+R_{d,k}^{\downarrow})$
regularize throughput and reserve. Each term is normalized as
$\widetilde J_j=J_j/S_j$. For $j=m,\Delta,f$, $S_j=J_j^0$ is the
corresponding uncontrolled value; $S_c=c_{\max}^{\mathrm{thr}}J_m^0$
and $S_R=K\sum_d\bar P_d$, with
$c_{\max}^{\mathrm{thr}}=\max_d c_d^{\mathrm{thr}}$. Thus, the weights
compare relative performance rather than quantities with different units.

\section{Case Studies}
\label{sec:case_studies}

\subsection{Experimental Settings}

\begin{figure}[t]
\centering
\includegraphics[width=0.9\columnwidth]{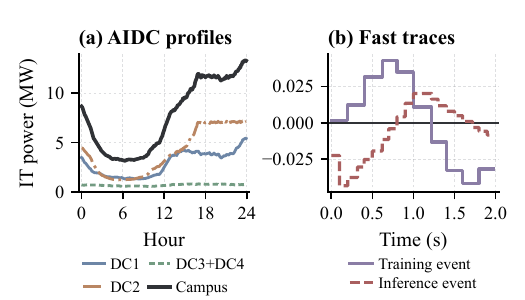}
\vspace{-1.em}
\caption{AIDC power profiles in the slow scheduling and fast response layers.}
\label{fig:data}
\end{figure}

The campus inputs use the NLR GenAI Power Profiles Dataset
\cite{Vercellino2026,Vercellino2026}. Four simulated
facility profiles provide the five-minute loads, and a strictly trailing
60-min mean defines $\bar L_{i,k}$. The fast-response analysis uses subsecond power deviations derived from measured two-node H100 training and online-inference traces. To construct campus-scale variations, the traces are replicated to represent 100 and 283 equivalent units, respectively, with 25\% of replicas synchronized. Each
trace is divided between two AIDCs in proportion to mean load, with a
one-source-sample offset, before aggregation at the PCC.
Fig.~\ref{fig:data} shows both time scales.

The shared HESS comprises a 3-MW/6-MWh Li-ion BESS and a 2-MW/0.5-MWh
flywheel. Their response parameters follow reported Li-ion performance
and the Beacon Power 450 XP specification
\cite{Petersen2019Performance,Beacon450XP}. The controller and device
steps are 100 and 20~ms, respectively, and the assumed campus pipeline
delay is 20~ms. Both devices operate within SOC bounds of 0.2--0.9 and
restore their initial SOC after the 24-h horizon. The BESS and flywheel
have initial SOCs of 0.55 and 0.60, charge/discharge efficiencies of
0.95/0.95 and 0.92/0.92, and reserve durations of 900 and 240~s,
respectively. Each fast window lasts 2~s, with $\tau_s=0.5$~s. The
weights $(\lambda_m,\lambda_{\Delta},\lambda_f,\lambda_c,\lambda_R)$
are $(1,0.25,1,0.02,0.005)$. The throughput coefficients are
$c_{\Esub}^{\mathrm{thr}}=8$ and
$c_{\Psub}^{\mathrm{thr}}=1$~USD/MWh. Six cases isolate capacity, topology, and response delay. C0 has no HESS;
C1 places a full HESS at each AIDC; C2 divides the C5 subsystem ratings
among four local HESSs. C3 uses one 5-MW/6.5-MWh BESS, matching C5 in total
power and energy. C4 uses the C5 ratings with instantaneous delivery,
whereas C5 includes the response in \eqref{eq:sampled_delivery}. Local cases retain device delays; C3 and C5 also include the campus pipeline.

Estimated installed CAPEX is calculated after dispatch as
$C^{\mathrm{cap}}=c_E^{\mathrm B}E^{\mathrm B}
+c_P^{\mathrm B}P^{\mathrm B}+c_P^{\mathrm F}P^{\mathrm F}
+N^{\mathrm{site}}c_0$, with energy in kWh, power in kW, and
$N^{\mathrm{site}}$ installations. The 2024 NLR ATB moderate commercial
battery model \cite{NLR2024ATBCommercial} gives
$c_E^{\mathrm B}=236.70$~USD/kWh,
$c_P^{\mathrm B}=904.61$~USD/kW, and
$c_0=207583.77$~USD per installation. The DOE/PNNL 0.25-h flywheel
estimate \cite{Mongird2019StorageCost} gives
$c_P^{\mathrm F}=3356.52$~USD/kW after conversion to 2022 USD using
BLS CPI-U \cite{BLS2026CPI}. The ATB coefficients are applied to the
study ratings; operating and replacement costs are excluded. Reported
throughput sums slow-layer charging and discharging energy. Conversion
loss sums
$(1-\eta_d^{\mathrm{ch}})p_{d,k}^{\mathrm{ch}}\Delta T+
(1/\eta_d^{\mathrm{dis}}-1)p_{d,k}^{\mathrm{dis}}\Delta T$
over subsystems, intervals, and installations.
\subsection{Multi-Timescale Regulation Performance}

\begin{figure}[t]
\centering
\includegraphics[width=0.8\columnwidth]{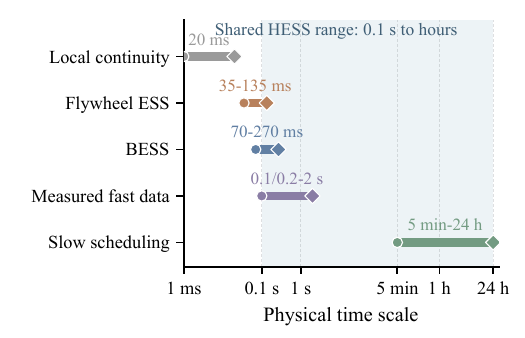}
\vspace{-1.5em}
\caption{Time scales of AIDC power variations and HESS service. Device bars
extend from response onset to 95\% response under the adopted parameters.}
\vspace{-1.em}
\label{fig:scope}
\end{figure}

The time scales in Fig.~\ref{fig:scope} explain the reason that scheduled capacity must be accompanied by an explicit delivery model. Under the model, the flywheel begins responding at 35~ms and reaches 95\% response at
135~ms but the corresponding BESS times are 70 and 270~ms. These delays are
comparable to the fast load variations. Consequently, a feasible five-minute schedule alone cannot establish
subsecond regulation capability. The slow scheduling layer must prepare power, energy,
and reserve, while \eqref{eq:sampled_delivery} determines how much response
reaches the PCC. Local UPSs retain responsibility for the earliest continuity
events, and the shared HESS addresses campus variations from approximately
0.1~s to hours.

\begin{figure}[t]
\centering
\includegraphics[width=0.85\columnwidth]{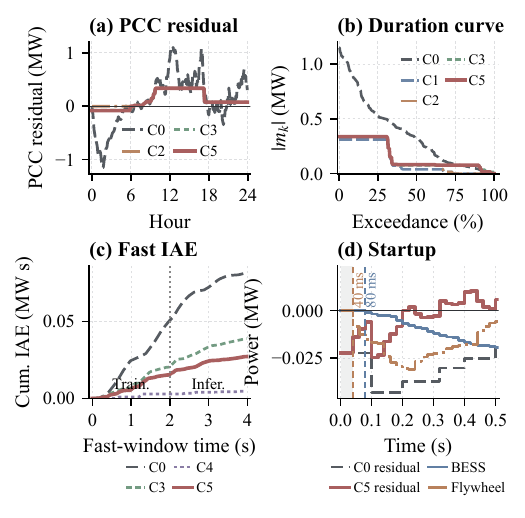}
\vspace{-1.5em}
\caption{PCC residuals and delivered fast response under the benchmark
configurations.}
\vspace{-1.5em}
\label{fig:performance}
\end{figure}

\begin{figure}[t]
\centering
\includegraphics[width=0.85\columnwidth]{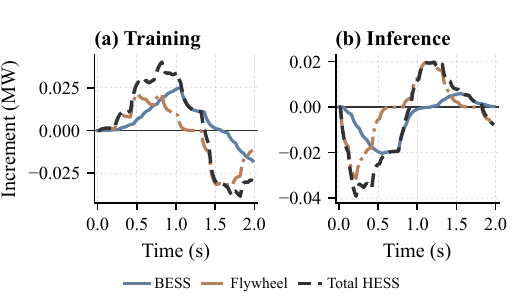}
\vspace{-1.5em}
\caption{BESS and flywheel responses in fast windows.}
\vspace{-1.5em}
\label{fig:mechanism}
\end{figure}

With these delivery limits enforced, the shared HESS reduces PCC IAE by
56.6\% and fast IAE by 66.7\% relative to C0
(Table~\ref{tab:key_results}). The flatter PCC trajectory and reduced upper
tail of the residual duration curve in Fig.~\ref{fig:performance}(a)--(b)
show how the slow schedule suppresses sustained departures from the campus
reference. At identical total power and energy ratings, C5 records a PCC IAE of 3.797~MWh versus 3.750~MWh for the BESS-only C3, while reducing fast
IAE by 30.1\%. The response decomposition in Fig.~\ref{fig:mechanism} explains this gain. The battery supplies smoother increments, and the flywheel responds more strongly to rapid changes in both training and inference windows, consistent with the causal split in \eqref{eq:causal_split}.

\begin{table}[t]
\caption{Operational Performance and Estimated Capital Cost}
\label{tab:key_results}
\centering
\renewcommand{\arraystretch}{1.05}
\setlength{\tabcolsep}{2pt}
\begin{tabular}{@{}lccccc@{}}
\toprule
Case &
\shortstack{PCC IAE\\(MWh)} &
\shortstack{Fast IAE\\(MW\,s)} &
\shortstack{Throughput\\(MWh)} &
\shortstack{Conversion\\loss (MWh)} &
\shortstack{Estimated CAPEX\\(million USD)}\\
\midrule
C0 & 8.748 & 0.0817 & 0      & 0     & 0\\
C1 & 2.851 & 0.0224 & 6.178  & 0.317 & 44.2186\\
C2 & 3.315 & 0.0224 & 11.396 & 0.781 & 11.6774\\
C3 & 3.750 & 0.0389 & 5.612  & 0.288 & 6.2692\\
C4 & 3.797 & 0.0047 & 5.603  & 0.299 & 11.0546\\
\textbf{C5} & 3.797 & 0.0272 & 5.603 & 0.299 & 11.0546\\
\bottomrule
\end{tabular}
\vspace{-1em}
\end{table}

C4 assumes instantaneous delivery of HESS commands and serves as an idealized case. Relative to the uncontrolled C0 case, C5 achieves
70.8\% of the fast-IAE reduction attained by C4 despite explicitly
accounting for these delays (Table~\ref{tab:key_results}). On the device grid, the first
flywheel and BESS increments occur at 40 and 80~ms, respectively
(Fig.~\ref{fig:performance}(d)). The initial disturbance remains unserved
until delivery begins. Despite this limitation, the accumulated fast
residual in Fig.~\ref{fig:performance}(c) remains below that of the
capacity-matched BESS in both workload windows. Thus, the hybrid advantage
persists after accounting for the response that can actually be delivered.

When the fast residual is scaled as
$\rho_w^{(\gamma)}(t)=\gamma\rho_w(t)$ with fixed ratings and delays,
C5 maintains lower fast IAE than the capacity-matched BESS C3 at every
level (Table~\ref{tab:fast_scaling}). The fast-energy excursions remain
at 3.47--10.42~Wh, supporting fast balancing around the inherited slow
operating point in \eqref{eq:fast_energy}.

\begin{table}[t]
\caption{Sensitivity to fast-residual scaling.}
\label{tab:fast_scaling}
\centering
\renewcommand{\arraystretch}{1.03}
\setlength{\tabcolsep}{1.6pt}
\begin{tabular}{@{}cccccc@{}}
\toprule
$\gamma$ &
\shortstack{Input peak\\(kW)} &
\shortstack{C3 IAE\\(MW\,s)} &
\shortstack{C5 IAE\\(MW\,s)} &
\shortstack{C5 RMS\\(MW)} &
\shortstack{C5 energy\\excursion (Wh)}\\
\midrule
0.5 & 21.6 & 0.0194 & 0.0136 & 0.0048 & 3.47\\
1.0 & 43.1 & 0.0389 & 0.0272 & 0.0095 & 6.95\\
1.5 & 64.7 & 0.0583 & 0.0408 & 0.0143 & 10.42\\
\bottomrule
\end{tabular}
\vspace{-1em}
\end{table}

\subsection{Operational and Economic Comparison}

Among all cases, although C1 achieves the lowest PCC IAE, it requires a full HESS at each AIDC, quadrupling the installed capacity and estimated CAPEX compared to the proposed method, C5. At matched total BESS and flywheel ratings, C2 achieves lower PCC and fast IAE than C5, but it does so relying on four local installations. Ultimately, the shared HESS, C5, retains 91.1\% of C2's PCC-IAE reduction and 91.9\% of its fast-IAE reduction relative to C0, while reducing throughput by 50.8\%, conversion loss by 61.7\%, and estimated CAPEX by $0.623$~million (Table~\ref{tab:key_results}).

The proposed C5 serves the aggregate campus residual through the PCC injection under
a common operating direction. Instead, C2 can charge a local HESS at one
AIDC while discharging another. With terminal SOC restored, this additional
cycling absorbs more of the negative PCC residual in
Fig.~\ref{fig:performance}(a) and increases the conversion loss in
\eqref{eq:slow_energy}. C2's additional 0.482~MWh conversion loss nearly
matches its 0.482~MWh lower PCC IAE.

Sharing also avoids repeating the installation cost at each AIDC. Because
the subsystem ratings are matched, the C2--C5 investment difference comes
from the ATB fixed term applied to four installations instead of one.
Estimated CAPEX falls from 11.6774 to 11.0546~million USD, saving
approximately \$0.623~million before assigning any monetary value to the
lower conversion loss. Although C3 is cheaper,
its higher fast IAE exposes the cost of omitting the faster subsystem.
The proposed model combines the lower storage use and installation
cost of shared operation with the subsecond response of hybrid storage.
Its regulation performance is established with the scheduled energy and
reserve limits and the delay of the single shared injection explicitly
represented.
\section{Conclusion}
\label{sec:conclusion}

Recognizing that AIDC campuses impose both sustained and subsecond power variations on the PCC, this paper proposes a multi-timescale model for a physically shared HESS. The model captures the electrical pooling and the shared PCC injection, linking scheduled power, energy state, and reserve capacity to subsecond delivered power while accounting for device and pipeline delays. In the NLR-based case studies,
C5 reduced PCC IAE by 56.6\% and fast IAE by 66.7\% relative to C0 baseline, and
achieved 70.8\% of the fast-IAE reduction available in the ideal case where commanded power is delivered immediately (C4). It also reduced fast IAE by 30.1\% relative to the capacity-matched BESS, C3. Although the capacity-matched local HESSs in C2 achieved lower IAE,
C5 retained more than 90\% of C2's reductions relative to C0 while halving
throughput, reducing conversion loss, and lowering estimated CAPEX by
\$0.623~million. These results show that the proposed shared HESS can mitigate campus power variations across both time scales when delivered power and resource use are accounted for. Future work will consider independent AIDC and storage operators.

\bibliographystyle{IEEEtran}
\bibliography{references}

@book{IEA2025EnergyAI,
  author    = {{International Energy Agency}},
  title     = {Energy and {AI}},
  publisher = {IEA},
  address   = {Paris, France},
  year      = {2025},
  url       = {https://www.iea.org/reports/energy-and-ai}
}

@techreport{NERC2025LargeLoads,
  author      = {{North American Electric Reliability Corporation}},
  title       = {Characteristics and Risks of Emerging Large Loads},
  institution = {North American Electric Reliability Corporation},
  type        = {Large Loads Task Force White Paper},
  address     = {Atlanta, GA, USA},
  month       = jul,
  year        = {2025}
}

@article{Latif2025Power,
 author  = {Imran Latif and Alex C. Newkirk and Matthew R. Carbone and Arslan Munir and Yuewei Lin and Jonathan Koomey and Xi Yu and Zhihua Dong},
  title   = {Single-Node Power Demand During {AI} Training: Measurements on an 8-{GPU} {NVIDIA H100} System},
  journal = {IEEE Access},
  volume  = {13},
  pages   = {61740--61747},
  year    = {2025}
}

@misc{Vercellino2026,
  author       = {Roberto Vercellino and Jared Willard and Gustavo Campos and Weslley da Silva Pereira and Olivia Hull and Matthew Selensky and Juliane Mueller},
  title        = {Measurement of Generative {AI} Workload Power Profiles for Whole-Facility Data Center Infrastructure Planning},
  howpublished = {\emph{arXiv preprint arXiv:2604.07345}},
  year         = {2026}
}

@article{Takci2025Flexibility,
  author  = {Mehmet T{\"u}rker Takci and Meysam Qadrdan and Jon Summers and Jonas Gustafsson},
  title   = {Data Centres as a Source of Flexibility for Power Systems},
  journal = {Energy Reports},
  volume  = {13},
  pages   = {3661--3671},
  year    = {2025},
  doi     = {10.1016/j.egyr.2025.03.020}
}

@article{Zheng2017HESS,
  author  = {Wenli Zheng and Kai Ma and Xiaorui Wang},
  title   = {Hybrid Energy Storage with Supercapacitor for Cost-Efficient Data Center Power Shaving and Capping},
  journal = {IEEE Transactions on Parallel and Distributed Systems},
  volume  = {28},
  number  = {4},
  pages   = {1105--1118},
  year    = {2017},
  doi     = {10.1109/TPDS.2016.2607715}
}

@article{Wang2023Hierarchical,
  author  = {Kaifeng Wang and Lin Ye and Shihui Yang and Zhanfeng Deng and Jieying Song and Zhuo Li and Yongning Zhao},
  title   = {A Hierarchical Dispatch Strategy of Hybrid Energy Storage System in Internet Data Center with Model Predictive Control},
  journal = {Applied Energy},
  volume  = {331},
  pages   = {120414},
  year    = {2023},
  doi     = {10.1016/j.apenergy.2022.120414}
}

@misc{You2026SourceSide,
  author       = {Haiyang You and Chengwei Lou and Jin Yang},
  title        = {Source Side Mitigation of {AI} Datacenter Power Fluctuations with a Hybrid Energy Storage System and Residual Differentiable Predictive Control},
  howpublished = {\emph{arXiv preprint arXiv:2606.04869}},
  year         = {2026}
}

@article{Han2023Shared,
  author  = {Ouzhu Han and Tao Ding and Xiaosheng Zhang and Chenggang Mu and Xinran He and Hongji Zhang and Wenhao Jia and Zhoujun Ma},
  title   = {A Shared Energy Storage Business Model for Data Center Clusters Considering Renewable Energy Uncertainties},
  journal = {Renewable Energy},
  volume  = {202},
  pages   = {1273--1290},
  year    = {2023},
  doi     = {10.1016/j.renene.2022.12.013}
}

@techreport{DOE2024Latency,
  author      = {{U.S. Department of Energy}},
  title       = {Latency Implications for Grid Communications},
  institution = {Office of Electricity, U.S. Department of Energy},
  type        = {SPaRC Series White Paper},
  address     = {Washington, DC, USA},
  month       = jan,
  year        = {2024}
}

@article{Petersen2019Performance,
  author = {L. Petersen and F. Iov and G. C. Tarnowski and V. Gevorgian and P. Koralewicz and D.-I. Stroe},
  title = {Validating Performance Models for Hybrid Power Plant Control Assessment},
  journal = {Energies},
  volume = {12},
  number = {22},
  pages = {4330},
  year = {2019}
}

@manual{Beacon450XP,
  author       = {{Beacon Power}},
  title        = {Beacon Power {450 XP} Flywheel Product Data Sheet},
  organization = {Beacon Power},
  year         = {2014}
}

@misc{NLR2024ATBCommercial,
  author       = {{National Renewable Energy Laboratory}},
  title        = {Commercial Battery Storage},
  howpublished = {2024 Annual Technology Baseline},
  year         = {2024},
  url          = {https://atb.nlr.gov/electricity/2024/commercial_battery_storage}
}

@techreport{Mongird2019StorageCost,
  author      = {K. Mongird and V. Fotedar and V. Viswanathan and V. Koritarov and P. Balducci and B. Hadjerioua and J. Alam},
  title       = {Energy Storage Technology and Cost Characterization Report},
  institution = {Pacific Northwest National Laboratory},
  number      = {PNNL-28866},
  month       = jul,
  year        = {2019}
}

@misc{BLS2026CPI,
  author = {{U.S. Bureau of Labor Statistics}},
  title  = {Summary of Annual and Semi-Annual Indexes},
  year   = {2026},
  url    = {https://www.bls.gov/regions/mid-atlantic/data/consumerpriceindexannualandsemiannual_table.htm}
}

@article{Bian2024Shared,
  author  = {Yifan Bian and Lirong Xie and Jiahao Ye and Lan Ma},
  title   = {A New Shared Energy Storage Business Model for Data Center Clusters Considering Energy Storage Degradation},
  journal = {Renewable Energy},
  volume  = {225},
  pages   = {120283},
  year    = {2024}

}

@article{Zhu2021Distributed,
  author  = {Hailing Zhu and Khmaies Ouahada},
  title   = {A distributed real-time control algorithm for energy storage sharing},
  journal = {Energy and Buildings},
  volume  = {230},
  pages   = {110478},
  year    = {2021}
}

@misc{Ko2025Mitigation,
  author       = {Min-Seung Ko and Jae Woong Shim and Hao Zhu},
  title        = {Mitigation of Datacenter Demand Ramping and Fluctuation using Hybrid {ESS} and Supercapacitor},
  howpublished = {\emph{arXiv preprint arXiv:2512.08076}},
  year         = {2025}
}

\end{document}